\documentstyle[preprint,aps,eqsecnum,floats]{revtex}

\newcommand{\beq}{\begin{equation}}
\newcommand{\eeq}{\end{equation}}
\newcommand{\beqa}{\begin{eqnarray}}
\newcommand{\eeqa}{\end{eqnarray}}

\newcommand{\gsim}{\mbox{${~\raise.15em\hbox{$>$}\kern-.85em
          \lower.35em\hbox{$\sim$}~}$}}
\newcommand{\lsim}{\mbox{${~\raise.15em\hbox{$<$}\kern-.85em
          \lower.35em\hbox{$\sim$}~}$}}

\newcommand{\Rbs}{\mbox{${\scriptstyle \not \!\!\; R}$}}

\def\npb#1{Nucl.\ Phys.\ {\bf B #1}}
\def\plb#1{Phys.\ Lett.\ {\bf B #1}}
\def\prd#1{Phys.\ Rev.\ {\bf D #1}}
\def\prl#1{Phys.\ Rev.\ Lett. {\bf #1}}

\def\gsim{\ \rlap{\raise 3pt \hbox{$>$}}{\lower 3pt \hbox{$\sim$}}\ }
\def\lsim{\ \rlap{\raise 3pt \hbox{$<$}}{\lower 3pt \hbox{$\sim$}}\ }

\begin{document}

\draft

{\tighten
\preprint{\vbox{\hbox{WIS-98/9/May-DPP}}}

\title{
Effects of R-parity Violating Couplings on CP
Asymmetries in Neutral $B$ Decays }

\author{Dafne Guetta\ }

\footnotetext{\footnotesize E-mail address:
guetta@bo.infn.it }

\address{
\vbox{\vskip 0.truecm}
  Dipartimento di Fisica \\
Universit\'a degli Studi di Bologna \\
Via Irnerio 46, I-{\it 40126}  \  Bologna, \ Italy}

\maketitle

\begin{abstract}%
A detailed analysis of the effects of
supersymmetric models without R-parity on various
CP asymmetries in neutral $B$ decays is given.
We concentrate on models with Abelian
horizontal symmetries that allow us to estimate the
order of magnitude of the new effects.
We focus on channels where the standard model gives
clean predictions: $ B_{d}\rightarrow\psi K_{S} $
and  $B_{d}\rightarrow\phi K_{S}$.
In the presence of new physics the two
asymmetries can  have a value different
from $\sin2\beta.$ Moreover,
they can  be different from each
other.

\end{abstract}

} 

\bigskip
\leftline{PACS number(s): 11.30.Er, 12.60.Jv, 13.20.He}

\newpage

{\tighten

\section{Introduction}
In the next few years several experiments on
$B$-physics will  take place (BaBar, BELLE, HERA B,
CLEO, RUN II at FNAL)\cite{esp}. They will provide
a test of many predictions of the standard model
(SM). Large part of the experiments will concentrate on
CP violation measurements. In this paper we study
the effects of the Minimal Supersymmetric
Standard  Model (MSSM)
without R-parity (with conserved baryon number)
on specific CP asymmetries.
These asymmetries can be significantly altered
from their SM values if there are important
new contributions to $ B_{d}-\bar{B}_{d} $ mixing
and/or to the decay amplitudes.

In the SM both the
$b$-quark decays and the $ B_{d}-\bar{B}_{d} $
mixing are determined by combinations of the CKM
matrix elements \cite{cpasim}. The asymmetries
measure the relative phases between these
combinations. Since the phases are related to
angles of the Unitarity Triangle (UT) through the CKM
unitarity, a measurement of the asymmetries
determines these angles.

Conversely, the SM predictions for the CP asymmetries in
neutral $B$ decays into certain CP eigenstates are
 determined by the values of the three angles
of the UT \cite{cpasim}. So, their
measurements will test these SM predictions and
consequently provide a  probe for physics beyond
the SM \cite{Quinn}.

Our work is an extension of previous studies done
on how R-parity violating models can affect CP
violation measurements \cite{Yuval,kaplan,Abel}.
We give  estimates of what the
models can do. In the spirit of
\cite{Dafne} we estimate the order of magnitude of
these new effects by embedding SUSY without
R-parity in models  with Abelian horizontal
symmetries.

The paper is organized as follows.
In sections II and III we give the general formalism
needed to study the CP asymmetries. In section IV we
briefly review SUSY
 models without R-parity.
In section V we study the effects of
R-parity violating couplings on the decay
amplitudes. We consider two CP asymmetries that
have equal values in the SM, but can be different
in R-parity violating models. The contribution to
the mixing is the same for all $ B_{d} $ decays,
so it cancels in the difference of the two  phases
that can be extracted from the asymmetries. We
estimate the maximal possible effects and analyze
whether  they are large enough to be clearly
signaled out.
In section VI we consider the contribution of the
R-parity violating couplings to the mixing and how
this affects the CP asymmetries.
In section VII
we estimate the effects of the R-parity violating
couplings in the
framework of models where the magnitude of the
fermion masses and CKM mixing angles is explained
by assuming some horizontal $U(1)$ symmetry.
Finally  section VIII contains our conclusions.

\section{General formalism for the definition of
CP asymmetry}

In this section we recall the relevant formulae for
the decay of neutral $B$ mesons into CP eigenstates
\cite{cpasim}.
The time dependent CP asymmetry is defined as:

\beq
\label{asicp}
a_{f_{CP}}(t) \equiv \frac{\Gamma [B^{0}_{phys}
(t)\rightarrow f_{CP}]-\Gamma [\bar{B}^{0}_{phys}
(t)\rightarrow f_{CP}]}{\Gamma [B^{0}_{phys}
(t)\rightarrow f_{CP}]+\Gamma [\bar{B}^{0}_{phys}
(t)\rightarrow f_{CP}]} ,
\eeq
where $B^{0}_{phys}$ and $\bar{B}^{0}_{phys}$
are states tagged as pure flavour eigenstate
$B_{d}$ and $\bar{B}_{d}$ at the production.

Eq.(\ref{asicp}) can be written as:
\beq
\label{acp}
 a_{f_{CP}}(t) \equiv
a_{f_{CP}}^{\cos} \cos(\Delta M t) +a_{f_{CP}}^{\sin}\sin(\Delta M t),
\eeq
where
\beqa
a_{f_{CP}}^{\cos} &=& \frac{ (1-|\lambda|^{2})}{1+|\lambda|^{2}},
\nonumber \\
a_{f_{CP}}^{\sin} &=& -\frac{2 {\rm Im} \lambda}{1+|\lambda|^{2}}.
\eeqa
 $\Delta M $ is the mass difference
between the two physical states, and
\beq
\label{landa}
\lambda \ =\ \sqrt{\frac{M_{12}^{*}
-\frac{i}{2}
\Gamma_{12}^{*}}{M_{12}-\frac{i}{2}
\Gamma_{12}}}\,
\frac{\langle f_{CP}|{\cal{H}}|B_{d}\rangle}{\langle
f_{CP}|{\cal{H}}|\bar{B}_{d}\rangle}\ =\
\frac{q}{p}\frac{\bar{A}}{A},
\eeq
 where $ M_{12} $ and
$\Gamma_{12}$ are the non diagonal
elements of the mass matrix {\bf M} and
of the matrix ${\bf \Gamma}$ which describes the
exponential decay of the system.
Since  $
\Gamma_{12}/M_{12}\simeq O(1\%)
\ll 1, $   we have
\beq \label{phim}
\left(\frac{q}{p}\right)^2\, =\, \frac{M_{12}^{*}
-\frac{i}{2}
\Gamma_{12}^{*}}{M_{12}-\frac{i}{2}
\Gamma_{12}}\,\simeq\, \frac{M_{12}^{*}}{M_{12}}
\,=\,\exp(-4i\phi_{M}),
\eeq
where $\phi_{M}  $ is the  $B-\bar{B}
$ mixing  phase.

The quantity $ {\rm Im}\lambda $ that can be
extracted from $a_{f_{CP}}(t)$ is directly related
to CKM matrix elements in the SM.
For mixing in the $B_{d} $ system, we
have $ M_{12}
\propto (V_{tb}V_{td}^{*})^{2} $ and consequently,
\beq
\left(\frac{q}{p}\right)_{B_{d}} =\frac{V_{tb}^{*}V_{td}}{
V_{tb}V_{td}^{*}}.
\eeq

As we mentioned in the introduction
CP asymmetries in decays to CP eigenstates,
$B_{d}\rightarrow f_{CP}, $ provide a way to
measure the three angles of the UT, defined by \cite{cpasim}
\beq
\alpha \equiv
\arg\left(-\frac{V_{td}V_{tb}^{*}}{
V_{ub}^{*}V_{ud}}\right)
\,\,\,
\beta \equiv
\arg\left(-\frac{V_{cd}V_{cb}^{*}}{
V_{tb}^{*}V_{td}}\right)
\,\,\,
\gamma \equiv
\arg\left(-\frac{V_{ud}V_{ub}^{*}}{
V_{cb}^{*}V_{cd}}\right).
\eeq

For example, in  the SM, we can measure $ \sin2\beta
$ in the decay $B_{d}\rightarrow \psi K_{S} $  where we have
$
\bar{A}/A
=[(V_{cb}V_{cs}^{*})/(V_{cb}^{*}V_{cs})]
[(V_{cs}V_{cd}^{*})/( V_{cs}^{*}V_{cd})],$ thus
\beqa
\lambda(B\rightarrow \psi K_{S}) &=&
\left(\frac{V_{td}V_{tb}^{*}}{ V_{td}^{*}V_{tb}}\right)
\left(\frac{V_{cb}V_{cs}^{*}}{ V_{cb}^{*}V_{cs}}\right)
\left(\frac{V_{cs}V_{cd}^{*}}{ V_{cs}^{*}V_{cd}}\right)
\nonumber \\
\rightarrow {\rm Im}\lambda &=& -\sin(2\beta).
\eeqa

\section{CP asymmetries in neutral B decays in
the presence of new physics}

We study CP asymmetries in neutral $B$ decays into final CP eigenstates.
In general, a decay amplitude has contributions from several diagrams
$A_i$ with different weak phases $\phi_i$ and different strong phases
$\delta_i$ \cite{cpasim}. If, for example,
there are two dominant contributions,
then the amplitudes $A$ for $B_{d}\rightarrow f_{CP}$ and
$\bar A$ for $\bar B_{d}\rightarrow f_{CP}$ are described by
\beqa
\label{amp}
A &=& A_{1}e^{i\phi_{1}}e^{i\delta_{1}}+
A_{2}e^{i\phi_{2}}e^{i\delta_{2}} \nonumber \\
\bar{A} &=& A_{1}e^{-i\phi_{1}}e^{i\delta_{1}}+
A_{2}e^{-i\phi_{2}}e^{i\delta_{2}}.
\eeqa
If $A_{2}/A_{1}\ =\ 0$ or $\phi_{1}\ =\ \phi_{2},$
then the CP asymmetry $a_{f_{CP}}$ cleanly
measures the CP violating quantity $\sin 2
(\phi_{M} +\phi_{1}),$ where $\phi_{M}$ is the
weak phase of the $B_{d}-\bar B_{d}$ mixing amplitude.
We focus on cases where within the SM
this is indeed the situation.

In the SM, the CP asymmetries in the decay modes
$b\rightarrow c\bar{c} s\,\, (B_{d}\rightarrow\psi
K_{S}),\,b\rightarrow c\bar{c} d\,\,
(B_{d}\rightarrow D^{+} D^{-} ),\,\, b\rightarrow
c\bar{u} d\,\, (B_{d}\rightarrow D^{0} \rho) $ and
 $ b\rightarrow s\bar{s} s\,\, (B_{d}\rightarrow\phi
K_{S})$ all measure the angle $\phi_M+\phi_1=\beta$ in $ B_{d}
$ decays even if they depend on different CKM factors.

{} From the theoretical point of view the cleanest mode
is  $b\rightarrow c\bar{c}s,$   for which the
penguin contribution, represented
by  $A_{2},$ has the same phase as the tree level
one, $A_{1},$ hence $\phi_{1}=\phi_{2}.$
In this case the theoretical
uncertainty for this channel is practically $\delta\phi_{SM} =0.$
This channel is the best mode also from the experimental point of view
because it has a relatively large branching ratio of $O(10^{-4})$
and can be identified by clean signals:
$\psi\rightarrow l^{+} l^{-},\,\, K_{S}
\rightarrow \pi^{+}\pi^{-} $
 \cite{esp,psi}.
 For these reasons this mode should be the reference mode
to which all the other measurements are compared.
Other clean modes are $b\rightarrow c\bar{u} d $
and $ b\rightarrow s\bar{s} s $ which have
$\delta\phi_{SM}<0.1 $ \cite{Yuval,isidori}, for
which the relevant exclusive branching ratios are
of $O(10^{-5}).$ Clean tests of the SM
aim  to see if the asymmetries for these decay
modes are consistent with the allowed range for
$\sin2\beta$ and, moreover, whether they are equal
to each other.

Among the channels given above, the most
sensitive to the effect of
the R-parity violating   models are
 the ones that
involve the second and the third family.
Therefore we concentrate on these two  modes:
$b\rightarrow c\bar{c} s\,\, (B_{d}\rightarrow\psi
K_{S})$ and
 $ b\rightarrow s\bar{s} s\,\, (B_{d}\rightarrow\phi
K_{S}).$

In the presence of new physics
a non-negligible $A_{2}$ (with a non-trivial weak phase $\phi_2$)
could arise for both channels. In addition, there could
be a new contribution to the mixing amplitude. Consequently, the
asymmetries may measure angles that are different from $\beta$ and,
moreover, different from each other.

 For simplicity we consider
 $\delta_{1}=\delta_{2} $ for both channels,
considering the fact that if they are not equal
we have direct CP violation and our analysis
can be generalized.
It follows that the ratio between the conjugate amplitudes is
a pure phase, $ \bar{A}/A
=e^{-2i\phi_{D}}. $ From eq.(\ref{landa}) and eq.(\ref{phim}), we have:
\beq
\lambda = e^{-2i(\phi_{M} +\phi_{D})},
\eeq
which implies that $ |\lambda |=1 $ and consequently
$a_{f_{CP}}^{\cos} = 0. $

In this case we can parameterize the new physics
effects in $a_{f_{CP}}$ by writing
\beq
\lambda = e^{2i(\phi_{0}+\delta\phi)},
\,\,\,
 a_{f_{CP}}(t) = -\sin[2(\phi_{0}+\delta\phi)]
\sin(\Delta M t) ,\eeq
where $\phi_{0}$ is the phase predicted at leading
order in the SM,
\beq a^{SM}_{f_{CP}}(t) \approx -\sin[2\phi_{0}]
\sin(\Delta M t) \eeq
and $\delta\phi$ is the correction due to new
physics (and subleading SM contributions, if they
are not negligible).

The R-parity violating couplings can contribute to
both the $ B_{d}-\bar{B}_{d}$ mixing amplitude and the
decay amplitude of a particular process. The
effect in the mixing will translate into
$\phi_{M}^{SM}\rightarrow
\phi_{M}^{SM}+\delta\phi_{M}^{\Rbs} $ and this shift
 is the same
for all $B_{d}$ decays. Although it changes the
value of the asymmetry with  respect to the SM, it
does not change the pattern predicted by the SM,
which is that the CP asymmetries  in
$ B_{d}\rightarrow\phi K_{S}$ and $ B_{d}\rightarrow\psi K_{S}$
are equal. The effect in the decay,
$\phi_{D}^{SM}\rightarrow
\phi_{D}^{SM}+\delta\phi_{D}^{\Rbs}, $ depends on the specific
process.
 For the two channels we mentioned before
 we consider the difference between the angles
$\phi(B_{d}\rightarrow f), $
obtained from the asymmetry measurement in the $
B_{d}\rightarrow f $ decay,
that has to be
zero in the SM up to theoretical uncertainties \cite{isidori},

\beq \label{dif}
| \phi(B_{d}\rightarrow\phi K_{S})-
\phi(B_{d}\rightarrow\psi K_{S}) | < 0.1 .\eeq
This equation is likely
to hold if there is not a large enhancement of the matrix
element of the $b\rightarrow u\bar{u} s$
operator between the $B_{d}$ initial state and the
$\phi K_{S}$ final state.
In \cite{isidori} the authors use $SU(3)$ relations
to suggest some experimental tests that can constrain
the size of this effect.
These tests
are likely to be done by the time that the CP
measurements will start.

If eq.(\ref{dif}) holds, it makes sense  to estimate the difference
 in the framework of
 SUSY models
without R-parity and to check if it can be larger than the SM
uncertainties.

\section{Susy without R-parity}
In this section we present a short introduction to
SUSY without R-parity. In contrast to the SM, in
the Supersymmetric version of the SM, the general
Lorentz and gauge invariant Lagrangian does not
have the accidental symmetries of baryon number
($B$) and lepton number ($L$). The most general
$L$ and $B$  violating superpotential is given by
\beq
W_{\Rbs} = \lambda_{ijk}L_{i}L_{j}\bar{l}_{k}+
\lambda^{\prime}_{ijk} L_{i}Q_{j}\bar{d}_{k}+
\lambda^{\prime\prime}\bar{u}_{i}\bar{d}_{j}\bar{d}_{k}
+\mu_{i} L_{i}\phi_{u};
\eeq
where
 $ i,j,k=1,2,3 $ are generation indices.
$ L_{i} (Q_{i})$  are the lepton (quark) $
SU(2)_{L} $ doublet superfields,
$\bar{l}_{j}(\bar{d}_{j},\, \bar{u}_{j})$  are the
charged leptons ( down and up quark ) $SU(2)_{L} $
singlet superfields and $\phi_{u}$ is the up Higgs
superfield. The bilinear terms are relevant for the
$\nu$-masses \cite{Enrico,Carlos}, but they are
not important for the present analysis  so we will
neglect them in the following. The first two terms
are $L$-violating and the third is $B$-violating.

The combination of $L$ and  $B$
violating operators will lead to rapid proton
decay. Then an extra symmetry is required to
enforce nucleon stability. Often an ad-hoc
symmetry, called R-parity, is imposed to keep $L$
and $B$ symmetries intact. This symmetry assigns a
charge $ R=(-1)^{3B+L+2S} $ to each particle,
where S is the particle's spin. Other possible
choices are to impose only $B$ or $L$.

For our analysis we consider  SUSY extensions of
the SM with highly suppressed $B$-violation (
$\lambda^{\prime\prime} =0 ),$ but without
R-parity \cite{susynoR} and without lepton number
\cite{banks} ($\lambda^{\prime},\,\,\lambda \neq
0$), which represent interesting alternatives to
the MSSM
\cite{MSSM}. The $\lambda^{\prime}$ terms give
rise to new contributions to $B$ decays.

Non-leptonic $B$-decays are caused by b-quark
transitions of the type $ b\rightarrow q_{1}
\bar{q}_{2} q_{3}, $ with
$ q_{1}\in \,\{d,s\} $ and $
q_{2},q_{3}\in\,\{u,d,c,s\}.$ The R-parity
violating effective terms relevant for these kind
of decays arise from the  slepton mediated tree level
diagrams. They  are of the form: $
(\lambda^{\prime}
\lambda^{\prime}/M^{2}_{(\tilde{l}\,{\rm or}\,\tilde{\nu})}
)\,(\bar{q}_{(L,R)}b_{(R,L)})(\bar{q}_{(R,L)}q_{(L,R)}),$
 where $M_{\tilde{l}}$ and $M_{\tilde{\nu}}$ are
the masses of the intermediate charged slepton and
sneutrino. In the following analysis we neglect
contributions of additional new physics operators
which  will affect the asymmetries at the loop
level.

Several models that can explain the observed
fermion mass hierarchy, like supersymmetric models
with horizontal symmetries \cite{miriam,horizontal}, also
predict that R-parity violating couplings
involving only the third and the second
generations fields are the largest ones
\cite{banks}. This is why we focus on
$b\rightarrow c\bar{c} s$ and $ b\rightarrow
s\bar{s} s $ decays.

\section{R-parity violating contributions
to the decay amplitudes}

Referring to eq.(\ref{amp}),  we can write the
decay amplitude in the presence of R-parity
violating couplings as:
\beq
A=A_{SM}+A_{\Rbs} = A_{SM}\Delta^{\Rbs}_{D},
\eeq
with
\beq
\Delta^{\Rbs}_{D}=1+r_{D} e^{i\theta_{D}},
\eeq
where $r_{D} = |A_{\Rbs}/A_{SM}|$ and
$\theta_{D}=\arg(A_{\Rbs}/A_{SM}).$
Note that $r_{D}$ is often constrained
phenomenologically while  $\theta_{D}$ is not.
 Consequently the phase $ \phi_{D} $ in the amplitude
 becomes $ \phi_{D}=\phi_{D}^{SM}+\delta\phi_{D} $
  where
  \beq
\delta\phi_{D} = \arg(\Delta^{\Rbs}_{D}) =
\arctan\left(\frac{r_{D} \sin\theta_{D}}{1+r_{D}
\cos\theta_{D}} \right).
\eeq

{}From this equation we   see  that if $ r_{D} \ll 1
$ then  $ \delta \phi_{D} \leq r_{D}. $  If however $r_{D}
\geq 1,$ then $\delta\phi_{D}$ could take any
value, because it depends strongly on
$\theta_{D}.$

 Since $r_{D} $ plays an important role in our discussion,
 we next estimate its order of magnitude for the two processes.

\subsection{Estimate of
$ r_{D}(B_{d}\rightarrow\phi K_{S}) $} The
amplitude for this process is given by:
\beq
A(B_{d}\rightarrow\phi K_{S})\
= \
\langle\phi K_{S}| {\cal{H}}_{eff}
|B_{d}\rangle,
\eeq
where
\beq
{\cal{H}}_{eff}={\cal{H}}^{SM}_{eff}+{\cal{H}}^{\Rbs}_{eff}.
\eeq
The Standard Model effective Hamiltonian relevant
for this channel is:
\beq {\cal{H}}^{SM}_{eff} = -
\frac{G_{F}}{\sqrt{2}}(V_{ts}^{*} V_{tb})
\sum_{k=3}^{10} Q_{k}c_{k}(\mu), \eeq
where $ c_{k}(\mu) $ are the Wilson coefficients,
which are functions of the scale $\mu$, with
$\mu=O(m_{b})$ being the relevant scale for $b$
decays. The operators $Q_{3}..Q_{6}$ are QCD
penguin operators and $Q_{7}..Q_{10}$ are electroweak penguin
operators\cite{Buras} .

The final $\phi$ requires that $Q_{k}$  lead to a
color-singlet term of the type
$\bar{s}\gamma^{\mu} s$. When Fierz-transforming
to obtain the required structure, some terms will
be multiplied by $1/N_{c}$ ($N_{c}$ is the number
of colors).

Our calculation uses the method of Deshpande and
He \cite{Deshpande}. In particular, we use
factorization to obtain the amplitude for
exclusive processes.
Using factorization introduces an unknown error
of order one in the final result.
We do not worry about uncertainties of this size because
we are trying only to get order of magnitude estimates.
In the SM we have:
\beq
A^{SM}(B_{d}\rightarrow\phi K_{S})=
-\frac{G_{F}}{\sqrt{2}}(V_{ts}^{*} V_{tb}) C
\langle K_{S}|\bar{s}\gamma^{\mu}(1-\gamma_{5})b|B_{d}\rangle
\langle \phi|\bar{s}\gamma^{\mu}s|0\rangle .
\eeq
C is the following combination of Wilson coefficients:
\beqa
C &=& (1+\frac{1}{N_{c}})\,c_{3}+ (1+\frac{1}{N_{c}})\,c_{4}+c_{5}
+\frac{1}{N_{c}}\,c_{6}\nonumber \\
&-&\frac{1}{2}(c_{7}+\frac{1}{N_{c}}\,c_{8} +
(1+\frac{1}{N_{c}})\,c_{9}+(1+\frac{1}{N_{c}})\,c_{10}).
\eeqa
Using  the results of ref.\cite{Buras}, we  can
give an estimate of C in the  leading
logarithmic approximation (LO), $ C\simeq 2\times
10^{-2}. $

For $b\rightarrow s\bar{s} s$ transitions the R-parity
violating contributions come from the sneutrino
mediated diagrams which give rise
 to these  new terms:
\beqa
c_{1}^{\Rbs}\,Q_{1}^{\Rbs} &=& \frac{1}{M^{2}_{\tilde{\nu}}}
(\lambda^{\prime *}_{i23} \lambda^{\prime}_{i22})
(\bar{s}^{\alpha}_{L} b_{R}^{\alpha})
(\bar{s}^{\beta}_{R} s_{L}^{\beta})
\nonumber \\
c_{2}^{\Rbs}\,Q_{2}^{\Rbs} &=& \frac{1}{M^{2}_{\tilde{\nu}}}
(\lambda^{\prime }_{i32} \lambda^{\prime
*}_{i22}) (\bar{s}^{\alpha}_{R} b_{L}^{\alpha})
(\bar{s}^{\beta}_{L} s_{R}^{\beta}).
\eeqa

The Fierz transformation introduces color
suppression factors:
\beqa
Q_{1}^{\Rbs}&\Longrightarrow &
\frac{1}{2}\frac{1}{N_{c}}(\bar{s}_{L}\gamma^{\mu}
 s_{L})(\bar{s}_{R}\gamma^{\mu} b_{R})
 \nonumber \\
Q_{2}^{\Rbs}& \Longrightarrow &
\frac{1}{2}\frac{1}{N_{c}}(\bar{s}_{R}\gamma^{\mu}
 s_{R})(\bar{s}_{L}\gamma^{\mu} b_{L}).
\eeqa

For ${\cal{H}}_{eff}^{\Rbs}$ we find:
\beqa
{\cal{H}}_{eff}^{\Rbs}&=&\frac{1}
{M^{2}_{\tilde{\nu}}}\frac{1}{N_{c}}\frac{1}{8}\left[
(\lambda^{\prime *}_{i23} \lambda^{\prime}_{i22})
(\bar{s}\gamma^{\mu}(1-\gamma_{5})s)
(\bar{s}\gamma^{\mu}(1+\gamma_{5})b) \right.
\nonumber \\
&+&\left.(\lambda^{\prime }_{i32} \lambda^{\prime *}_{i22})
(\bar{s}\gamma^{\mu}(1+\gamma_{5})s)
(\bar{s}\gamma^{\mu}(1-\gamma_{5})b) \right]
.\eeqa

Since
\beq
\langle K_{S}|\bar{s}\gamma^{\mu}(1-\gamma_{5})b|B_{d}\rangle =
\langle K_{S}|\bar{s}\gamma^{\mu}(1+\gamma_{5})b|B_{d}\rangle =
\langle K_{S}|\bar{s}\gamma^{\mu}b|B_{d}\rangle ,
\eeq
and
\beq
\langle \phi|\bar{s}\gamma^{\mu}(1-\gamma_{5})s|0\rangle =
\langle \phi|\bar{s}\gamma^{\mu}(1+\gamma_{5})s|0\rangle =
\langle \phi|\bar{s}\gamma^{\mu}s|0\rangle,
\eeq
the R-parity contribution to the amplitude modifies the SM
amplitude by
an overall coefficient:
\beqa
A(B_{d}\rightarrow\phi K_{S}) &=&
A^{SM}(B_{d}\rightarrow\phi K_{S})+
A^{\Rbs}(B_{d}\rightarrow\phi K_{S})\nonumber \\
&=&\left[ -\frac{G_{F}}{\sqrt{2}}(V_{ts}^{*}
V_{tb}) C +\frac{1}
{M^{2}_{\tilde{\nu}}}\frac{1}{N_{c}}
\frac{1}{8} (\lambda^{\prime *}_{i23}
\lambda^{\prime}_{i22}
 + \lambda^{\prime }_{i32}\lambda^{\prime*}_{i22})\right]
 \langle \phi|\bar{s}\gamma^{\mu}s|0\rangle  \langle K_{S}
|\bar{s}\gamma^{\mu}b|B_{d}\rangle
.\eeqa

We are now in a position to give $
r_{D}(B_{d}\rightarrow\phi K_S) $:
\beq
r_{D}(B_{d}\rightarrow\phi K_S)
=\frac{|\frac{1}{M^{2}_{\tilde{\nu}}}\frac{1}{N_{c}}
\frac{1}{8}(\lambda^{\prime *}_{i23} \lambda^{\prime}_{i22}
 + \lambda^{\prime }_{i32}\lambda^{\prime*}_{i22})|}
 {|\frac{g^{2}}{8 M^{2}_{W}}(V_{ts}^{*}V_{tb}) C |}.
\eeq
Putting $|V_{ts}^{*} V_{tb}|=4\times 10^{-2},$
$N_{c}=3,$ $ C\simeq 2\times 10^{-2} $ we estimate
\beq
r_{D} \simeq 8 \times 10^{2}\, |\lambda^{\prime
*}_{i23} \lambda^{\prime}_{i22}
 + \lambda^{\prime }_{i32}\lambda^{\prime*}_{i22}|\,
\left(\frac{M_{W}}{M_{\tilde{\nu}}} \right)^{2}.
\eeq
The bound $ BR(b\rightarrow X_{s}\nu\bar{\nu})
\leq 7.7\times 10^{-4} $ \cite{aleph} implies the following
limits on products of R-parity violating couplings
\cite{Nardi}:
\beq \label{costraints}
\lambda^{\prime }_{i32}\lambda^{\prime*}_{i22}
< 1.5\times 10^{-3}\
\left(\frac{m_{\tilde{d}_{L}}}{100 GeV}\right)^{2}.
\eeq
For the product of the other couplings we use the
limits found in \cite{agasha}
\beq
\lambda^{\prime *}_{i23} \lambda^{\prime}_{i22}
< 1.4\times 10^{-4}\
\left(\frac{m_{\tilde{d}_{R}}}{100 GeV}\right)^{2}.
\eeq
Using these limits and
 taking  $m_{\tilde{d}}\sim M_{\tilde{\nu}} $ we get
\beq r_{D} (B_{d}\rightarrow\phi K_S)\leq 0.8,
\eeq
and so
\beq
\delta\phi(B_{d}\rightarrow\phi K_S) \lsim 0.8 .
\eeq
\subsection{Estimate of
$ r_{D}(B_{d}\rightarrow\psi K_{S}) $}

In \cite{kaplan} the estimate of
the maximal value of $r_{D}$
was done considering  the matrix
elements of the R-parity violating and SM
operators of the same order.
We obtain the same result
following the  steps of the previous analysis for

\beq A(B_{d}\rightarrow\psi K_{S})
\ = \ \langle \psi K_{S}|
{\cal{H}}_{eff} |B_{d}\rangle .
\eeq

The relevant SM Hamiltonian is:

\beq {\cal{H}}^{SM}_{eff} =
\frac{G_{F}}{\sqrt{2}}(V_{cb}^{*} V_{cs})
\sum_{k=1}^{10} Q_{k}c_{k}(\mu). \eeq

We find:
\beq
A^{SM}(B_{d}\rightarrow\psi
K_{S})=\frac{G_{F}}{\sqrt{2}}(V_{cb}^{*} V_{cs})
C^{\prime}
\langle K_{S}|\bar{s}\gamma^{\mu}(1-\gamma_{5})b|B_{d}\rangle
\langle \psi|\bar{c}\gamma^{\mu}c|0\rangle ,
\eeq
where $ C^{\prime}=c_{1}+\frac{1}{N_{c}}\,c_{2} .$ In the LO , $
C^{\prime}\simeq 8\times 10^{-2} $ \cite{Buras}.

For $b\rightarrow c\bar{c} s$ transitions the R-parity
violating contributions come from the charged slepton
mediated diagrams which give rise
 to a  new term which Fierz transforms to:

\beqa
c_{1}^{\Rbs}\,Q_{1}^{\Rbs} &=&
\frac{1}{M^{2}_{\tilde{l}}}
\frac{1}{N_{c}}\frac{1}{8}(\lambda^{\prime
*}_{i23} \lambda^{\prime}_{i22})
(\bar{c}_{L}\gamma^{\mu} c_{L})
(\bar{s}_{R}\gamma^{\mu} b_{R}).
\eeqa
The R-parity contribution to the amplitude is to modify the
overall  coefficient:
\beqa
A(B_{d}\rightarrow\psi K_{S}) &=&
A^{SM}(B_{d}\rightarrow\psi K_{S})+
A^{\Rbs}(B_{d}\rightarrow\psi K_{S})\nonumber \\
&=&\left[ \frac{G_{F}}{\sqrt{2}}(V_{cb}^{*}
V_{cs}) C^{\prime} +\frac{1}
{M^{2}_{\tilde{l}}}\frac{1}{N_{c}}
\frac{1}{8}
(\lambda^{\prime *}_{i23} \lambda^{\prime}_{i22})
 \right]
\nonumber \\
& & \langle \psi|\bar{c}\gamma^{\mu}c|0\rangle
\langle K_{S}|\bar{s}\gamma^{\mu}b|B\rangle
,\eeqa
leading to
\beq
r_{D} (B_{d}\rightarrow\psi K_S)=\frac{|\frac{1}
{M^{2}_{\tilde{l}}}\frac{1}{N_{c}}
\frac{1}{8}
(\lambda^{\prime *}_{i23}
\lambda^{\prime}_{i22})|}
 {|\frac{g^{2}}{8 M^{2}_{W}}(V_{cb}^{*}
V_{cs}) C^{\prime} |.}
  \simeq 2 \times 10^{2} \,
|\lambda^{\prime *}_{i23}
\lambda^{\prime}_{i22}|\,
\left(\frac{M_{W}}{M_{\tilde{l}}}
\right)^{2}.
\eeq
Putting the same limits as before for
$\lambda^{\prime *}_{i23}
\lambda^{\prime}_{i22}, $  and taking $ M_{\tilde{l}}\simeq
m_{\tilde{d}},$ we find $ r_{D} \leq 0.02 $
which is in agreement with what found in \cite{kaplan}.
Then we have
\beq
\delta\phi(B_{d}\rightarrow\psi K_S) \lsim 0.02 .
\eeq
We conclude  that in the presence of R-parity
violating couplings it is possible to obtain:
\beq
|\phi(B_{d}\rightarrow\phi K_S) -
\phi(B_{d}\rightarrow\psi K_S)|
\sim O(1).
\eeq
Comparing this result  with eq.(\ref{dif}) we
learn that  new physics in decay amplitudes could
lead to deviation from the pattern of CP violation
in $ B_{d} $ decays  larger than the SM uncertainties.

\section{Effects of R-parity violating couplings on
$ B_{d}-\bar{B_{d}} $ mixing }

In order to give a complete analysis of how R parity
violating couplings affect the CP asymmetries, we
have to consider the effect
on $ B_{d}-\bar{B_{d}} $ mixing.

The mixing amplitude can be written as
\beq
 M_{12} =M_{12}^{SM}+M_{12}^{\Rbs} =M_{12}^{SM}\Delta_{M}^{\Rbs},
\eeq
where we define
\beq
\Delta_{M}^{\Rbs} = 1+  r_{M} e^{i\theta_{M}},
\eeq
with $ r_{M}=|M_{12}^{\Rbs}/M_{12}^{SM}| $
and $ \theta_{M} = \arg\left[
\lambda^{\prime *}_{i13}\lambda^{\prime}_{i31}/
 (V_{tb}^{*} V_{td})^{2} \right].$
The mixing angle becomes  $ \phi_{M}
=\phi^{SM}_{M}+ \delta\phi_{M}, $ where
\beq \label{argd}
 \delta\phi_{M}=\arg(\Delta_{M}^{\Rbs} )
=\arctan\left( \frac{r_{M}\sin\theta_{M}}
{1+r_{M}\cos\theta_{M}}\right).
\eeq

In order to  estimate  the contribution
to the mixing, we first study the phenomenological
constraints on
$r_{M}$ and then we estimate it in the
model with horizontal symmetry.

Note that the only experimental limits we have on the R-parity
 violating couplings that contribute to the mixing come
from the bound on $\Delta M _{B_{d}} $ itself.
Since the R-parity violating contribution
could saturate $\Delta M_{B} $ and the SM contribution
could be less than a half of it, it is clear that
$r_M > 1 $ is allowed.

{}From eq.(\ref{argd}) we see that,  if $r_{M}>1,$
$\delta\phi_{M}$ depends strongly on $\theta_{M}
,$
 which, in turn, depends on the new phases of
the $\lambda^{\prime}$ couplings. As there are no
experimental constraints on the value of these
phases, the asymmetries could have any value,
unlike the SM case where the phases are constrained,
 i.e. $0.3 \lsim \sin 2\beta \lsim 0.9 .$

Although we concluded that $r_{M}$ could be large, it
is still useful to give an explicit expression for it.
This is done in the next subsection.

\subsection*{Estimate of $ r_{M}(B_{d}\rightarrow f) $}

The mixing matrix element is given by:
\beq
|M_{12}| =\frac{1}{M_{B_{d}}} |\langle \bar{B}_{d} | {\cal{H}}_{eff}(\Delta B
=2) | B_{d}\rangle |,
\eeq
the SM contribution is \cite{Buras}
\beq
M_{12}^{SM} (B_{d})=
\frac{G_{F}^{2}}{12\pi^{2}} \eta_{B} M_{B_{d}} (B_{B_{d}} f^{2}_{B_{d}})
M_{W}^{2} S_{0}(x_{t}) (V_{tb}^{*} V_{td})^{2},
\eeq
where $\eta_{B} =0.55, $\cite{Jamin},  $x_{t}=(m_{t}/M_{W})^2$ and
 $ S_{0}(x_{t}) $ is given in \cite{Inami}.

R-parity
violating terms contribute to $ M_{12} $
at tree level  through sneutrino-exchange.
In fact, the effective Hamiltonian term for
$ B_{d}-\bar{B}_{d} $ is
\beq
{\cal{H}}_{eff}^{\Rbs} =  \frac{\lambda^{\prime
*}_{i13}\lambda^{\prime}_{i31}}
{M^{2}_{\tilde{\nu}}}  Q_{S}(\Delta B =2 )+h.c.
\eeq
 where
\beq
Q_{S}(\Delta B= 2 )= (\bar{d}_{L}b_{R})(\bar{d}_{R}b_{L})
\eeq
In the vacuum insertion approximation  \cite{Ecker}:
\beq
\langle \bar{B}_{d} | Q_{S}(\Delta B
=2) | B_{d}\rangle = \left(\frac{M_{B_{d}}^{2}}{(m_{b}+m_{d})^{2}}+
\frac{1}{6}\right)  f^{2}_{B_{d}} M^{2}_{B_{d}}.
\eeq
Then the R-parity violating mixing term is
\beq
M_{12}^{\Rbs} =\frac{\lambda^{\prime *}_{i13}\lambda^{\prime}_{i31}}
{M^{2}_{\tilde{\nu}}} \left(\frac{M_{B_{d}}^{2}}{(m_{b}+m_{d})^{2}}+
\frac{1}{6}\right)  f^{2}_{B_{d}} M_{B_{d}},
\eeq
leading to
\beq
r_{M} =\frac{|\frac{\lambda^{\prime *}_{i13}\lambda^{\prime}_{i31}}
{M^{2}_{\tilde{\nu}}} \,(\frac{M_{B_{d}}^{2}}{(m_{b}+m_{d})^{2}}+
\frac{1}{6})|}{|\frac{G_{F}^{2}}{12\pi^{2}} \eta_{B} B_{B_{d}}
M_{W}^{2} S_{0}(x_{t}) (V_{tb}^{*} V_{td})^{2}|}
\simeq  10^{8}\,\,
|\lambda^{\prime *}_{i13}\lambda^{\prime}_{i31}|
\,\left(\frac{100\,GeV}
{M_{\tilde{\nu}}}\right)^{2}.
\eeq

\section{ R-parity violation
                    in the framework of horizontal symmetries}
\vspace{0.1in}
In the previous sections we found the upper bounds
on $r_{D}(B_{d}\rightarrow f) $ from
phenomenological constraints and we gave the general
expression of $r_{M} $ as a function of the
R-parity violating couplings.
In this section we
estimate $ r_{D} $ and $r_{M}$ in the framework of
supersymmetric  models with Abelian horizontal
symmetries that have been thoroughly investigated
in \cite{miriam,horizontal}. These models explain
naturally the order of magnitude of the fermion
masses, the hierarchy among them and the CKM
angles. Assuming a horizontal $U(1)$ symmetry
with a small breaking parameter, $ \varepsilon\sim 0.2,$
allows us to estimate the size of the
$\lambda^{\prime}$ $L$
violating couplings relevant for our processes
 and to work out numerical
predictions for $ r_{D} $. Most of the
$L$-violating couplings are  suppressed with
respect to the corresponding Yukawa couplings $
Y_{ij}.$ They can be  estimated as
\cite{Dafne}

\beq  \label{lpnum}
\lambda^\prime_{kij} \sim  Y^d_{ij} \varepsilon^{H(L_k) - H(\Phi_d)},
\eeq
where $ H(\psi)$ is the charge assigned to the
field $\psi.$
Using the equations (2.2)-(2.4) of ref.\cite{miriam}
we obtain
\beq \label{lprime}
\lambda^\prime_{kij}
          \lsim \sqrt{G_F} \tan\beta \>
  (|V_{ij}|)^{{\rm sign}(j-i)} \>
m_{d_{j}} \>
\varepsilon^{H(L_k) - H(\Phi_d)}.
\eeq

The  equation (\ref{lprime})
shows that like the
lepton and down quark Yukawa couplings, the
$\lambda^\prime$ couplings increase with
$\tan\beta =
\langle \phi_{u} \rangle/\langle \phi_{d} \rangle
$.
\subsection{Estimate of $r_{D}$ for the two processes}
We consider two different
frameworks of the horizontal symmetries
that allow us to find
order of magnitude estimates for $r_{D}$
in these models.

\paragraph{}
In a framework where the neutrino
masses are suppressed only by the horizontal
symmetries \cite{banks} we can write
\beq
\varepsilon^{H(L_k) - H(\Phi_d)}\lsim
\sqrt{\frac{m_{\nu_{\tau}}}{m_{Z}}},
\eeq
therefore the constraints on the
$\lambda^{\prime}$
couplings come from the bound on the mass of the
$\tau$-neutrino, $m_{\nu_{\tau}}\lsim 18\,MeV \cite{neutrino},$
 obtained from direct experiments.

The suppression of the $\lambda^{\prime}$
couplings is then related to
fermion parameters in the following way:
\beqa \label{rd}
r_{D}(B_{d}\rightarrow \phi K_{S})  &\lsim& 2\times 10^{-3} \>
 G_F\> \tan^2\beta \>\frac{m_{s}^{2}}{|V_{cb}V_{ts}|}
 \nonumber \\
 r_{D}(B_{d}\rightarrow \psi K_{S})  &\lsim&
5\times 10^{-4} \> G_F\> \tan^2\beta \> m_{s}
\> m_{b}.
\eeqa
We fix $\tan\beta =1 $ and obtain the numerical results:
\beqa
r_{D}(B_{d}\rightarrow \phi K_{S})  &\lsim&
5\times 10^{-7} \nonumber \\
 r_{D}(B_{d}\rightarrow \psi K_{S})  &\lsim& 5\times 10^{-9},
\eeqa
leading to
\beq \label{case1}
|\phi(B_{d}\rightarrow \phi K_{S})-\phi(B_{d}\rightarrow \psi K_{S})| \lsim 5\times 10^{-7}.
\eeq
Then  in this framework we obtain unobservable deviations from the
SM predictions.
\paragraph{}
Within a framework in which the neutrino masses
are not suppressed by the horizontal symmetries but by
alternative mechanisms, (like the alignment of
the $\mu$ and $B$ terms \cite{Enrico})
 we relate our couplings to the ones
involved in processes where the R-parity violating terms
induce flavour changing neutral currents,
 like $B-\bar{B}$ and $K-\bar{K}$ mixing, which
are the most constrained.
 In this framework we get the maximal values allowed
for $\lambda^{\prime}$ in agreement with all
present constraints.

The strongest bound is the one on the product of the
couplings involved in $K-\bar{K}$
mixing \cite{mixing}
\beq
\lambda^{\prime}_{i21} \lambda^{\prime}_{i12} \leq 5\times 10^{-9}.
\eeq
Then writing our couplings in terms of these ones
and of the fermion parameters we obtain for $r_{D}$
of the two processes:
\beqa
r_{D}(B_{d}\rightarrow \phi K_{S})  &\sim& 20\>
\left(\frac{|V_{us}|}{|V_{cd} V_{cb} V_{ts}|}\>
\frac{m_{s}}{m_{d}}\right)\>
| \lambda^{\prime}_{i21} \lambda^{\prime}_{i12}| \nonumber \\
 r_{D}(B_{d}\rightarrow \psi K_{S})  &\sim& 5\>
\left(\frac{|V_{us}|}{|V_{cd}|}\>\frac{m_{b}}{m_{d}}\right)\>
|\lambda^{\prime}_{i21} \lambda^{\prime}_{i12}|.
\eeqa
Putting the numerical values and the
constraints on the couplings, we have:
\beqa
r_{D}(B_{d}\rightarrow \phi K_{S})  &\lsim& 1\times 10^{-3}  \nonumber \\
 r_{D}(B_{d}\rightarrow \psi K_{S})  &\lsim& 1\times 10^{-5},
\eeqa
leading to
\beq \label{case2}
|\phi(B_{d}\rightarrow \phi K_{S})-\phi(B_{d}\rightarrow \psi K_{S})|
 \lsim 1\times 10^{-3}.
\eeq
We see that also in this case we do not obtain sizeable effects.

{} From  eq.(\ref{case1}) and
eq.(\ref{case2}) we conclude that this
 type of R-parity violating SUSY models are
unlikely to be signaled out through the
comparison of these two CP asymmetries.
However in \cite{Dafne} it was found that
a signature of these models could
indeed be detected in
rare $B$
decays in the third family leptons.
Therefore we can conclude that  rare $B$ decays are more
sensitive to this
kind of effects than CP asymmetries.

\subsection{Estimate of $r_{M}$}

For the mixing, in the framework $a$ of the horizontal symmetries
 we have:
\beq
r_{M}\lsim   10^{4}\,
G_F \tan^2\beta \> \frac{|V_{ub}|}{|V_{tb}|} \>m_{b}\> m_{d}.
\eeq
Inserting the numerical values we obtain:
\beq
r_{M}\lsim 4\times 10^{-3},
\eeq
and the correction to the mixing phase is limited to be:
\beq
\delta\phi_{M} \lsim 4\times 10^{-3}.
\eeq

We conclude also that this effect is too small to
be signaled out through the measurement of
the CP asymmetries.

In the framework $b$ we find for $r_{M}$
\beq
r_{M}\sim 10^{8}\>\left( \frac{|V_{ub}|}
{|V_{ts}V_{cd}|}\>\frac{m_{b}}{m_{s}}\right)\>
|\lambda^{\prime}_{i21} \lambda^{\prime}_{i12}|.
\eeq
Putting the numerical values and the
constraint on the product of the
couplings, we obtain
\beq r_{M}\sim 1 . \eeq
Then $\delta\phi_{M}$ depends
strongly on $\theta_{M}$,
\beq
\delta\phi_{M} \sim \frac{\theta_{M}}{2},
\eeq
 on which we do not have any
constrain. This implyes that unlike the
SM case
the CP asymmetries can assume any value.
Even if in this case the new physics can
affect strongly the mixing phase, again
a measurement of the CP asymmetries  does not allow us to
identify this particular kind of new physics.

\section{Conclusions}
We have studied new CP violating effects
that can arise in
models without R-parity and without lepton number.
We focused on
 the CP asymmetries for the two channels
$ B_{d}\rightarrow\psi
K_{S} $ and $B_{d}\rightarrow\phi K_{S}$
because these observables are experimentally
accessible, have small theoretical
uncertainties and are sensitive to this
type of new physics.

In the SM these decays measure directly the  phase
$\beta, $ up to small corrections due to the presence of subleading
 contributions of order $10\% $ in $B_{d}\rightarrow\phi K_{S}.$

In general we have found that   $B_{d}\rightarrow\phi
K_{S}$ and the $ B-\bar{B} $ mixing may be
strongly affected, while the amplitude of the mode
$ B_{d}\rightarrow\psi K_{S} $ remains almost the same that in the SM.
The result of the
R-parity violating couplings could be a deviation
of both the CP asymmetries from $\sin 2\beta$ and
in a sizeable difference between them. In SUSY
models without R-parity and with Abelian
horizontal symmetries we have considered two possible
situations.
When $\nu$-masses are suppressed only by the
horizontal symmetry, we have found
negligible effects in the amplitudes and a weak
effect in the mixing.
In models where the $\mu$ and the $B$ terms
are aligned so that $\nu$ masses arise only from loop
effects (or nonrenormalizable terms),
the effect of the R-parity violating couplings could
be larger.
However,
also in this case the effect on the
amplitudes remains small
while the effect on the mixing can be sizeable.
We conclude that the CP asymmetries
are not sensitive to this kind of new physics.
Other observables like branching ratios
of rare $B$ decays into  third family leptons
are better suited for the study of
this kind of effects \cite{Dafne}.

\vspace{1.5cm}

\acknowledgements
I thank  Yossi Nir for suggesting this work,
Roberto Casalbuoni, Yuval Grossman for relevant
contributions to the paper and Adam Falk, Alex
Kagan, Enrico Nardi for useful comments on the
manuscript. I also acknowledge the Particle
Theory Group of the Weizmann Institute of Science,
where this research was carried out, for the
hospitality and for the pleasant working
atmosphere.

}  
\end{document}